\documentclass[12pt]{article}

\usepackage{latexsym}

\usepackage[font=small,labelfont=bf]{caption}

\usepackage{graphics}
\usepackage{graphicx}
\usepackage{epstopdf}
\usepackage{amssymb}
\usepackage{tabularx}
\usepackage{caption}
\usepackage{subcaption}
\usepackage{hyperref}
\usepackage{tabularx}
\usepackage{gensymb}

\begin{document}


\title {Image of the thin accretion disk around compact objects in the Einstein-Gauss-Bonnet gravity}

\author{
Galin Gyulchev$^{1}$\footnote{E-mail: \texttt{gyulchev@phys.uni-sofia.bg}},  \, Petya Nedkova$^{1}$\footnote{E-mail: \texttt{pnedkova@phys.uni-sofia.bg}},  \, Tsvetan Vetsov$^{1}$\footnote{E-mail: \texttt{vetsov@phys.uni-sofia.bg}},\\ Stoytcho Yazadjiev$^{1,2}$\footnote{E-mail: \texttt{yazad@phys.uni-sofia.bg}}\\ \\
   {\footnotesize${}^{1}$ Faculty of Physics, Sofia University,}\\
  {\footnotesize    5 James Bourchier Boulevard, Sofia~1164, Bulgaria }\\
    {\footnotesize${}^{2}$ Institute of Mathematics and Informatics,}\\
{\footnotesize Bulgarian Academy of Sciences, Acad. G. Bonchev 8, } \\
  {\footnotesize  Sofia 1113, Bulgaria}}
\date{}
\maketitle

\begin{abstract}
We study the optical appearance of a thin accretion disk around compact objects within the Einstein-Gauss-Bonnet gravity. Considering static spherically symmetric black holes and naked singularities we search for characteristic signatures which can arise in the observable images due to the modification of general relativity. While the images of the Gauss-Bonnet black holes closely resemble the Schwarzschild black hole, naked singularities possess a distinctive feature. A series of bright rings are formed in the central part of the images with observable radiation $10^3$ times larger than the rest of the flux making them observationally significant. We elucidate the physical mechanism, which causes the appearance of the central rings, showing that the image is determined by the light ring structure of the spacetime. In a certain region of the parametric space the Gauss-Bonnet naked singularities possess a stable and an unstable light ring. In addition the gravitational field becomes repulsive in a certain neighbourhood of the singularity. This combination of features leads to the formation of the central rings implying that the effect is not specific for the Einstein-Gauss-Bonnet gravity but would also appear for any other compact object with the same characteristics of the photon dynamics.
\end{abstract}

\section{Introduction}

Recently a new static spherically symmetric black hole was proposed which generalizes the Schwarzschild black hole when quantum corrections are considered \cite{Cai:2010}-\cite{Cognola:2013}. It was derived as an exact solution to the semi-classical Einstein equations with Weyl anomaly of the energy-momentum tensor by restricting the trace anomaly only to the Gauss-Bonnet invariant. The new black holes possess a logarithmic term in their Bekenstein-Hawking entropy, which is consistent with the microscopic statistical interpretation of the black hole entropy in string theory or quantum gravity. In addition, they provide a resolution to the singularity problem, since the gravitational force becomes repulsive at very small distances from the compact object preventing the particles and light to access the singularity.

The solution was also obtained in a different context considering the classical four-dimensional (4D) Einstein-Guass-Bonnet equations \cite{Glavan:2020}. It was suggested that the Einstein-Gauss-Bonnet theory possesses a non-trivial limit to four spacetime dimensions which circumvents the Lovelock theorem and allows the contribution of the Gauss-Bonnet term to the local dynamics. While the proposed regularization procedure is not consistent for general gravitational fields \cite{Wen}-\cite{Fernandes}, it leads to correct predictions in a number of cases with high symmetries, such as spherically symmetric solutions for example.

Irrespective of the theoretical framework in which the solution is interpreted, the non-trivial contribution of the Gauss-Bonnet term is expected to have phenomenological impact. This triggered a range of works investigating different observational features such as the shadow \cite{Zeng}, the gravitational lensing \cite{Jin}, and the radiation from the accretion disk  \cite{Liu}. Static relativistic stars in the proposed  4D limit of  the Einstein-Gauss-Bonnet gravity were constructed in \cite{Doneva}, while the gravitational collapse of a spherical cloud of  dust  was studied in \cite{Malafarina}.

The purpose of this work is to investigate the optical appearance of four-dimensional compact objects with a Gauss-Bonnet term. Similar studies were initiated in order to search for the observational signatures in the electromagnetic spectrum of various types of compact objects in general relativity and the modified theories of gravity \cite{Nedkova:2019}-\cite{Kunz:2016}. We construct the observable images of the thin accretion disk around black holes and naked singularities belonging to the class of solutions obtained in \cite{Cai:2010}-\cite{Glavan:2020}.   While Gauss-Bonnet black holes appear very similar to the Schwarzschild black hole,  we observe significant distinctions in the case of naked singularities. The accretion disk possesses multiple images, which form a series of concentric bright rings in the central region of the primary disk image. Similar effect was obtained recently for the strongly naked Janis-Newman-Winicour singularity where its appearance was connected with the absence of a photon sphere \cite{Nedkova:2020}. However, in the case of the Gauss-Bonnet naked singularities the multiple images are related to a different feature of the photon dynamics, and they appear even when the solution possesses a photon sphere.  We elucidate the physical mechanism which leads to the formation of the image noting that similar processes may also take place in the case of other compact objects where multiple ringlike disk images are observed in the presence of a photon sphere \cite{Shaikh:2019}-\cite{Vincent:2020}.

The paper is organized as follows. In the next section we describe the exact solution representing static compact objects within Gauss-Bonnet theory and some of the characteristics of the geodesic motion in this spacetime, which are relevant for our studies. In section 3 we present the images of the thin accretion disk around Gauss-Bonnet black holes and weakly naked singularities as seen by a distant observer and discuss their properties. In section 4 we analyze the structure of the images revealing the mechanism of their formation. In the last section we summarize our results.

\section{Properties of the exact solution}

Static spherically symmetric compact objects within 4D Gauss-Bonnet gravity can be described by the metric \cite{Cai:2010}

\begin{eqnarray}\label{metric}
ds^2 &=& -f(r)dt^2 + \frac{1}{f(r)}dr^2 + r^2(d\theta^2 + \sin^2\theta \phi^2), \\
f(r) &=& 1 + \frac{r^2}{2\gamma}\left(1-\sqrt{1+ \frac{8\gamma M}{r^3}}\right). \nonumber
\end{eqnarray}
where $\gamma$ is a positive constant and $M$ is the $ADM$ mass. It was derived originally as an exact solution to the semi-classical Einstein equations with one loop quantum corrections

\begin{eqnarray}
R_{\alpha\beta} - \frac{1}{2}R g_{\alpha\beta} = 8\pi\langle T_{\alpha\beta}\rangle,
\end{eqnarray}
when we consider only the Gauss-Bonnet term in the general form of the trace anomaly of the effective energy-momentum tensor $\langle T_{\alpha\beta}\rangle$

\begin{eqnarray}
g^{\alpha\beta}T_{\alpha\beta} = -\gamma\left(R^2 - 4R_{\alpha\beta}R^{\alpha\beta} + R_{\alpha\beta\gamma\delta}R^{\alpha\beta\gamma\delta}\right),
\end{eqnarray}
where the coupling constant $\gamma$  depends on the degrees of freedom of the quantum fields. The metric can be alternatively derived by solving the classical D-dimensional Einstein-Gauss-Bonnet equations  and taking a non-trivial limit to four-dimensional spacetime \cite{Glavan:2020}.

The properties of the solution depend on the value of the dimensionless coupling constant $\hat\gamma=\gamma/M^2$. When $\hat\gamma$ belongs to the range $\hat\gamma \in [0,1]$ the solution describes a black hole with an inner and an outer horizon located at $r_{\pm}= M \pm \sqrt{M^2-\gamma}$, respectively. The value $\hat\gamma=1$ corresponds to the extremal black hole, while $\hat\gamma=0$ is the Schwarzschild limit. The solutions with $\hat\gamma >1$ are naked singularities. The Kretschmann invariant  diverges at the location of the singularity $r=0$, however in a slower rate than for the Schwarzschild solution since it behaves as $R_{\alpha\beta\gamma\delta}R^{\alpha\beta\gamma\delta}\sim \frac{1}{r^3}$ when approaching the singularity.

In this work we will consider black holes, i.e. solutions with $\hat\gamma \in (0,1]$ and naked singularities with coupling constant taking the range $1<\hat\gamma < 3\sqrt{3}/4$. In this region of the parametric space the solution describes weakly naked singularities, which  possess a photon sphere.\footnote{We use the classification of the naked singularities by means of their lensing properties to weakly and strongly naked, which was introduced by Virbhadra when studying  of the Janis-Newman-Winicour solution \cite{Virbhadra:2002}.} The photon dynamics is determined additionally by the presence of a stable photon ring located in the interior of the photon sphere. The location of the photon rings can be determined by obtaining the stationary points of the effective potential for the null geodesics

\begin{equation}
V^{ph}_{eff}= L^2\frac{f(r)}{r^2},
\end{equation}
where $L$ is the specific momentum of the photon. The radial coordinates of the photon rings correspond to the solutions of the system $V^{ph}_{eff}=0$ and $dV^{ph}_{eff}/dr=0$, which can be reduced to the following algebraic equation

\begin{eqnarray}
r^3+ 8M\gamma -9M^2r=0.
\end{eqnarray}

\begin{figure*}[t!]
    \centering
    \begin{subfigure}[t]{0.7\textwidth}
        \includegraphics[width=\textwidth]{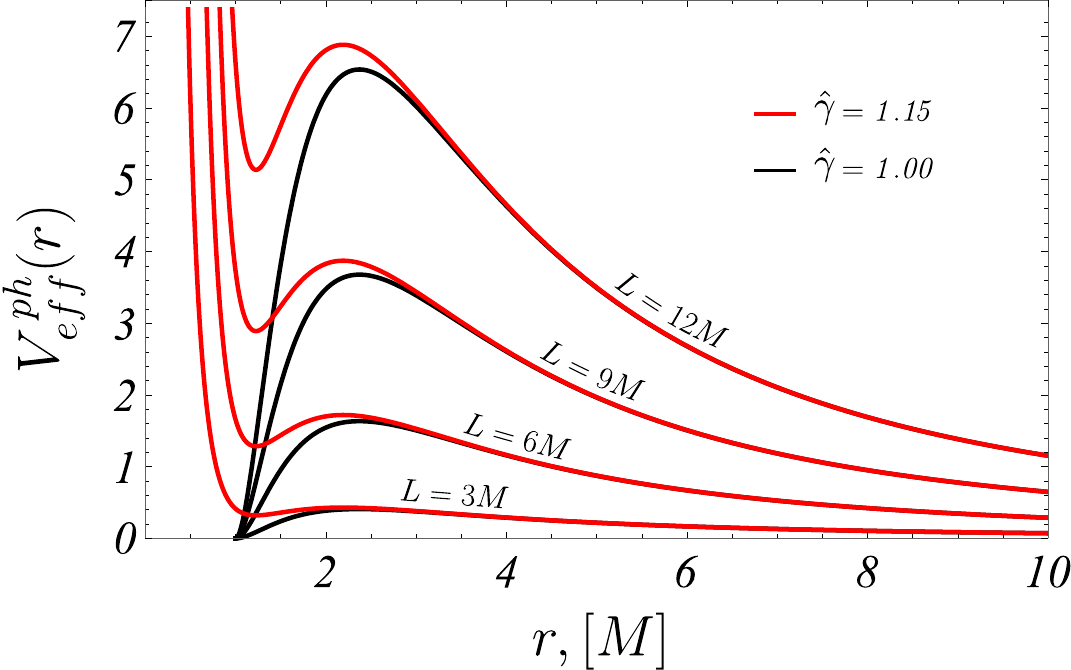}
           \end{subfigure}
           \caption{\label{fig:V_eff}\small Effective potential for the null geodesics for the Gauss-Bonnet black holes with $\hat\gamma=1$  and weakly naked singularities with $\hat\gamma=1.15$.}
\end{figure*}

On the other hand, the Gauss-Bonnet black holes possess only a photon sphere similar to the Schwarzschild black hole. In fig. $\ref{fig:V_eff}$ we illustrate the behavior of the effective potential for the two types of compact objects as the photon sphere corresponds to the maximum of the potential, while the stable light ring is located at the minimum.


The structure of the timelike circular orbits for the Gauss-Bonnet black holes  resembles the Schwarzschild black hole. The stable circular orbits extend from a certain marginally stable orbit to infinity. For weakly naked singularities the stable circular orbits are located in two disconnected region. There exists an inner annular region located between the inner limit of existence of timelike circular orbits $r_{in}$ and the stable photon ring, and an outer region extending from the marginally stable timelike orbit $r_{ms}$ to infinity. In the gap between the two regions we have the following behavior. Timelike circular orbits are not allowed between the stable and unstable photon rings, while after passing  the photon sphere they exist but remain unstable until reaching $r_{ms}$. The inner radius of existence of timelike circular orbits is given by the solution of the equation

\begin{eqnarray}
r^3 + 2M\gamma - r^3\sqrt{1+ \frac{8M\gamma}{r^3}} =0,
\end{eqnarray}
while the location of the marginally stable orbit corresponds to the inflection point of the effective potential $V_{eff}$ for the timelike geodesics satisfying $V_{eff}=0$, $dV_{eff}/dr=0$ and $d^2V_{eff}/dr^2=0$, where

\begin{equation}
V_{eff}= f(r)\left(1 +\frac{L^2}{r^2}\right).
\end{equation}
In the following analysis the position of the $r_{ms}$ is obtained numerically.

For our purposes the described characteristic features of the circular geodesic orbits are sufficient to determine the structure of the thin accretion disk for coupling constant in the range $0<\hat\gamma < 3\sqrt{3}/4$. A detailed qualitative analysis of the geodesic motion will be presented in a further work.

\section{Image of the thin accretion disk around Einstein-Gauss-Bonnet black holes and naked singularities}

We consider the image a thin accretion disk around Einstein-Gauss-Bonnet black holes and weakly naked singularities with $1<\hat\gamma < 3\sqrt{3}/4$. The disk is described by the Novikov-Thorne model \cite{Novikov}, \cite{Page:1974}, and consists of radiating particles moving on stable circular geodesics in the compact object's spacetime. We construct its image as seen by a distant observer by obtaining the optical appearance of the timelike circular orbits at spacetime infinity and evaluating the observable radiation flux.

The apparent shape of the timelike circular orbits is obtained numerically using a ray-tracing procedure. We apply additionally a semi-analytical scheme requiring only numerical integration, which is more convenient for the interpretation of the images. Both methods are described in detail in our previous works \cite{Nedkova:2019}-\cite{Nedkova:2020}. In essence, the trajectories of the photons emitted at a certain point on the circular orbits are obtained numerically up to some radial distance corresponding effectively to the spacetime infinity. Afterwards,  their projections on the observer's sky are visualized by means of two celestial angles $\alpha$ and $\beta$, which are related to the photon's 4-momentum as

\begin{eqnarray}\label{p_alpha}
&&p_\theta=\sqrt{g_{\theta\theta}}\sin\alpha,\qquad\qquad\,\,\, p_\phi  =\sqrt{g_{\phi\phi}}\sin\beta\,\cos\alpha, \nonumber \\[2mm]
&&p_r=\sqrt{g_{rr}}\cos\beta\,\cos\alpha.
\end{eqnarray}

Practically it is more convenient to use  $\alpha$ and $\beta$ as initial data and integrate the photon trajectory backwards to its emission point. We consider the full range of the celestial angles $\alpha\in[0,\pi]$ and $\beta\in[-\pi/2, \pi/2]$, and select those values which correspond to  null geodesics originating from the accretion disk, i.e. solutions to the geodesic equation passing through the equatorial plane at a radial coordinate with range of stability of the timelike circular orbits.  In this way we obtain the set of the celestial angles, which represent the image of the accretion disk on the observer's sky.

We associate further with each point of the image an observable flux  emitted by the accreting particles. The local flux radiated by a portion of the disk delimited by the radial coordinates $r_0$ and $r$ is evaluated according to the Novikov-Thorne model by the integral \cite{Novikov}, \cite{Page:1974}

\begin{equation}\label{F_r}
 F(r)=-\frac{\dot{M}_{0}}{4\pi \sqrt{-g^{(3)}}}\frac{\Omega
_{,r}}{(E-\Omega
L)^{2}}\int_{r_0}^{r}(E-\Omega
L)L_{,r}dr,
\end{equation}
where $r_0$ is corresponds to the edge of the disk,  $\dot{M}_{0}$ is the accretion rate, and $g^{(3)}$  is the determinant of the induced metric in the equatorial plane. We denote by $\Omega$, $E$ and $L$ the angular velocity, the energy and the angular momentum of the particles on the circular orbits. For a general static spherically symmetric metric in the form

\begin{equation}
ds^2=g_{tt}\,dt^2+g_{\phi\phi}\,d\phi^2+g_{rr}\,dr^2+g_{\theta\theta}\,d\theta^2\,,
\end{equation}
they are given by the expressions

\begin{eqnarray}
E&=&-\frac{g_{tt}}{\sqrt{-g_{tt}-g_{\phi\phi}\Omega^2}},    \label{rotE}  \\[2mm]
L&=&\frac{g_{\phi\phi}\Omega}{\sqrt{-g_{tt}-g_{\phi\phi}\Omega^2}},     \nonumber  \\[2mm]
\Omega&=&\frac{d\phi}{dt}=\sqrt{-\frac{g_{tt,r}}{g_{\phi\phi,r}}}.    \nonumber
\end{eqnarray}

In order to obtain the observable flux $F_{obs}$ at a given point of the celestial sphere we should take into account the gravitational redshift $z$. Thus, we obtain the relation

\begin{equation}\label{F_obs}
F_{obs} = \frac{F}{(1+z)^4},
\end{equation}
where the gravitational redshift  for a general spherically symmetric spacetime can be expressed by means of the metric functions and the impact parameter $b=L/E$ as \cite{Luminet:1979}

\begin{equation}
1+z=\frac{1+\Omega b}{\sqrt{ -g_{tt} - \Omega^2 g_{\phi\phi}}}.
\end{equation}

\begin{figure}[t!]
    		\setlength{\tabcolsep}{ 0 pt }{\small\tt
		           \includegraphics[width=1.\textwidth]{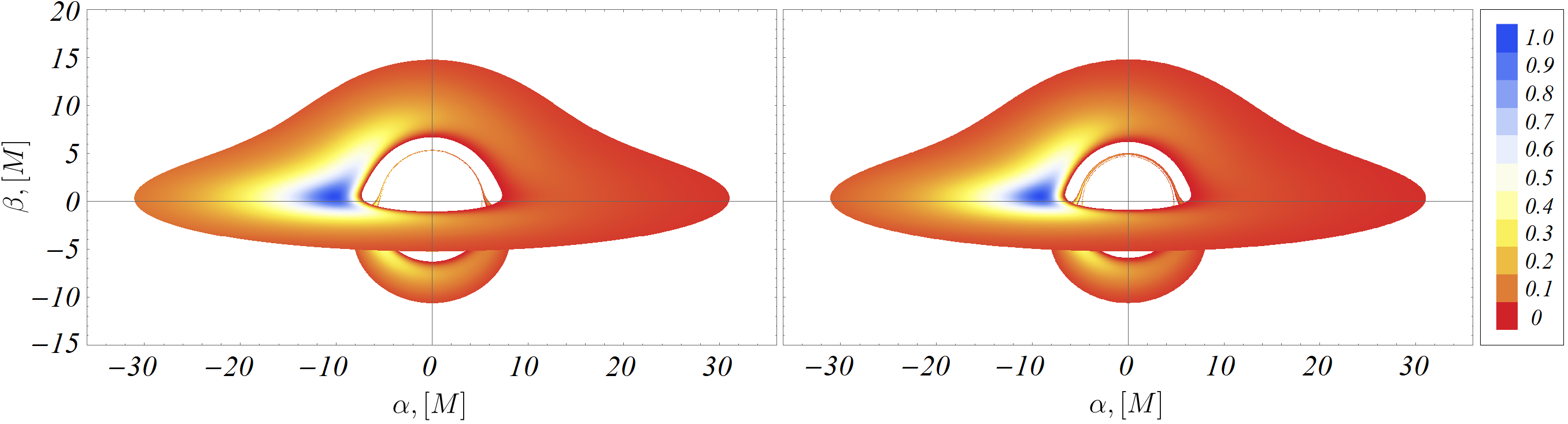} }
		   \caption{\label{fig:ColorDisk1}\small Apparent radiation flux for the Gauss-Bonnet black hole with $\hat\gamma=1$ (left), and for the Schwarzschild black hole (right). The observer is located at $r_{obs} = 5000 M$, and at the inclination angle $i=80^\circ$. }
\end{figure}

\begin{figure}[h]
	\setlength{\tabcolsep}{ 0 pt }{\small\tt
		
           \includegraphics[width=1.\textwidth]{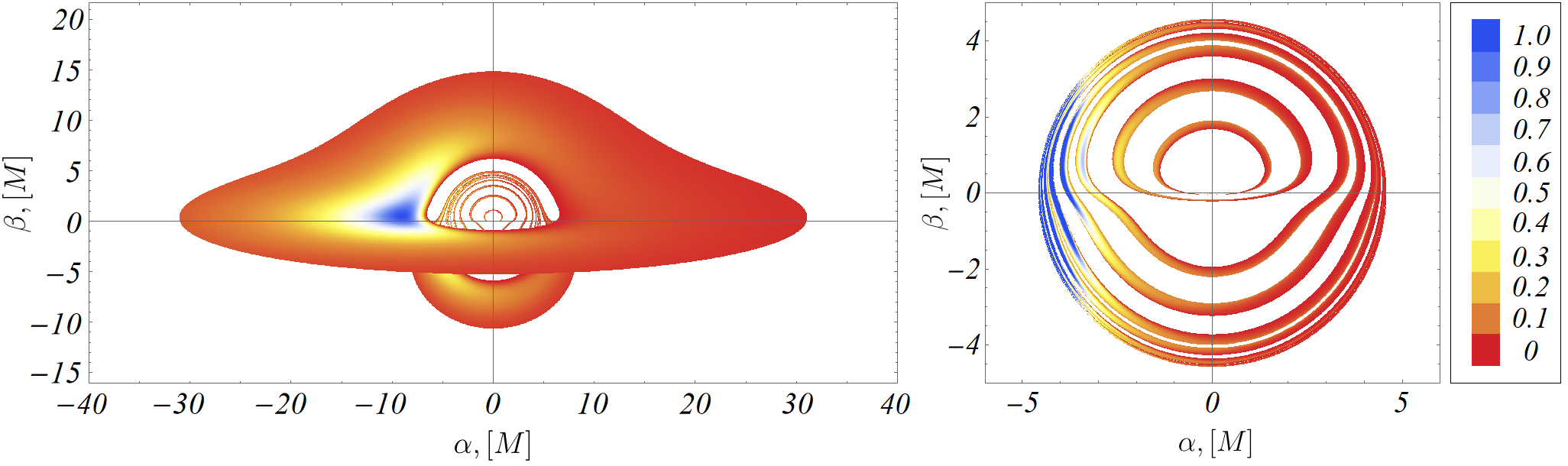}}\caption{\label{fig:ColorDisk2}\small Apparent radiation flux for the Gauss-Bonnet weakly naked singularity with $\hat\gamma=1.15$. The radiation from the outer disk (left) and the inner disk (right) are represented separately in order to make the image more transparent. The maximum of the observable radiation from the outer disk is $F^{out}_{max}= 6.07\times 10^{-5} M \dot{M}$, while for the inner disk the flux reaches the value $F^{in}_{max}= 4.94 \times 10^{-2} M \dot{M}$. The observer is located at $r_{obs} = 5000 M$, and at the inclination angle $i=80^\circ$. }
\end{figure}

We illustrate the qualitative behavior of the images for the Einstein-Gauss-Bonnet black holes and the weakly naked singularities by choosing a representative value of the coupling constant $\hat\gamma$ for each class of solutions. Their optical appearance is given in Figs. $\ref{fig:ColorDisk1}$ and $\ref{fig:ColorDisk2}$ representing the distribution of the apparent flux intensity as seen by a distant observer, where $\hat\gamma = 1$ for the black hole, and $\hat\gamma =1.15 $ for the naked singularity. The image of the Schwarzschild black is also given for comparison. The observable flux is normalized by its maximal value, and its range $F_{obs}/F^{max}_{obs}\in[0,1]$ is mapped continuously to the color spectrum from red to blue, as the highest values are depicted in blue. The effective infinity is assumed to correspond to the radial coordinate $r=5000 M$, and we choose an inclination angle $i=80^\circ$, since for larger inclination angles the relativistic effects are more pronounced.

The images of the Einstein-Gauss-Bonnet black holes resemble closely the  Schwarz- \\ schild black hole. Although we choose a Guass-Bonnet solution with a value of the coupling constant maximally deviating from the Schwarzschild black hole, only slight quantitative differences are present. The disk size is smaller in the Einstein-Gauss-Bonnet case and the peak of the observable radiation is slighly lower compared to the Schwarzschild black hole. Still, the observable flux distribution follows the same pattern for the two solutions with a similar characteristic location of the maximum of the radiation in the vicinity of the ISCO on the left-hand side of the image.

On the other hand for weakly naked singularities we observe clear qualitative distinctions. In the central region of the image a series of bright rings are formed which are absent in the case of the Schwarzschild or the Gauss-Bonnet black holes. For weakly naked singularities the stable circular orbits are distributed into two disconnected regions which form an inner and an outer accretion disk. The outer disk produces two types of images - a hat-like image similar to the Schwarzschild black hole, and a sequence of rings located in the interior part of the image of the marginally stable orbit. On the other hand the inner disk is observed only in the form of central ring images.

The ring images resulting from the two disconnected parts of the accretion disk radiate with different intensity. The observable flux of the rings resulting from the outer disk can be evaluated to be less than $30 \%$ of the apparent flux maximum from the outer disk. The maximum of the observable radiation from the outer disk $F^{out}_{max}$ is located on the hat-like image similar to the Schwarzschild and Gauss-Bonnet black holes.  In contrast the observable flux from the inner disk rings is with three orders of magnitude larger as its maximum $F^{in}_{max}$ is related to the radiation from the outer disk as $F^{in}_{max}/F^{out}_{max}\sim 10^{3}$. Thus, it represents the most pronounced intensity in the radiation from the whole accretion disk image.

In order to facilitate the interpretation of the image  we present the optical appearance of the outer and the inner disks separately in the left and right panels of Fig. $\ref{fig:ColorDisk2}$. The complete observable image of the accretion disk is composed by the superposition of the two images. For the chosen value of the coupling constant $\hat\gamma=1.15$ the outer accretion disk extends from the marginally stable orbit $r_{ms} = 5.08 M$ to infinity. The inner disk is located between the inner limit of existence of the timelike circular orbits $r_{in} = 1.05 M$ and the stable photon ring $r_{ph} = 1.23 M$.

A characteristic feature of the inner disk is that the emitted flux $F(r)$ decreases when the radius of the particle orbit $r$ moves towards its inner edge. This is contrary to the behavior of the flux for the outer disk, or the accretion disk for the Schwarzschild or Gauss-Bonnet black holes, which has a maximum in the vicinity of the ISCO. The reason is the opposite gradient of the angular velocity on the circular orbits in this region, which decreases in direction to the naked singularity becoming zero at the disk inner edge. Such anomaly can be associated with the repulsive action of the gravitational field in the vicinity of the singularity, which prevents the particles from falling in, thus screening the singularity.

\section{Image formation for the weakly naked singularities}

Using a semi-analytical scheme for constructing the optical appearance of the circular orbits we can study the process of the image formation for the weakly naked singularities. This procedure is described in detail in \cite{Nedkova:2019}-\cite{Nedkova:2020}, therefore here we online only the basic steps. The central quantity in the scheme is the evaluation of the variation of the azimuthal angle $\phi$ along the photon trajectory. For a general spherically symmetric metric in the form ($\ref{metric}$) it is given by the integral
\begin{equation}\label{phi_b}
\phi(b)= \int^{r_{obs}}_{r_{source}}{\frac{dr}{r^2\sqrt{\frac{1}{b^2} -\frac{f(r)}{r^2}}}},
\end{equation}
where $b=L/E$ is the impact parameter on the geodesic and the integration is performed between the photon's emission point $r_{source}$ and the location of the observer $r_{obs}$. Taking advantage of the spherical symmetry, the azimuthal angle can be expressed by means of the inclination angle $i$ and a properly defined celestial coordinate $\eta$ as \cite{Luminet:1979}, \cite{Muller:2009}

\begin{equation}\label{phi_eta}
\phi = -\arccos{ \frac{\sin\eta\tan i}{\sqrt{\sin^2\eta\tan^2 i + 1}}}.
\end{equation}

The equality of the two expressions $(\ref{phi_b})$ and $(\ref{phi_eta})$  determines the impact parameter of the photon trajectories which can be observed at a given inclination angle and a given celestial coordinate\footnote{For an observer located at the asymptotic infinity the celestial coordinate $\eta$ is related to the impact parameters $\alpha$ and $\beta$, which we used in the visualization of the disk images as $\eta = \arctan{\frac{\beta}{\alpha}}$.}. Considering the full range of the celestial coordinate $\eta\in[0,\pi]$ we get the impact parameters of all the photon trajectories emitted by a particular circular orbit with a radial coordinate $r=r_{source}$,  which can reach an observer located at $r=r_{obs}$ and inclination angle $i$. So far we considered only trajectories which reach the observer directly without revolving around the compact object. In general we can observe also trajectories performing an arbitrary large number of turns around the origin of the coordinate system. Such trajectories are taken into account by including in the equation an integer number $k$, parameterizing the number of half-loops around the coordinate origin. Direct trajectories with $k=0$ lead to the primary observable image, but we can observe also secondary images of higher order $k$. Thus, we obtain the basic equation

\begin{equation}\label{orbit_im}
\int^{r_{obs}}_{r_{source}}{\frac{dr}{r^2\sqrt{\frac{1}{b^2} -\frac{f(r)}{r^2}}}} = k\pi -\arccos{ \frac{\sin\eta\tan i}{\sqrt{\sin^2\eta\tan^2 i + 1}}}.
\end{equation}

We should further consider that some trajectories possess a  radial turning point. Then, the integral $(\ref{phi_b})$ is represented as a sum of two integrals including the turning point in the integration limits. The turning point $r_0$ is a function of the impact parameter $b$ determined by the largest root of the equation

\begin{eqnarray}\label{D_r}
b = \frac{r_0}{\sqrt{f(r)}}.
\end{eqnarray}

We can illustrate graphically the solutions of eq. ($\ref{orbit_im}$) by using the following argument. For every order $k$ the right-hand side of the equation determines a minimal and a maximal observable azimutal angle, which we denote by $\phi^{k}_{min}$ and $\phi^{k}_{max}$, respectively. They correspond to the minimal and maximal values of the celestial coordinate $\eta$. Then, for every order $k$ we define an observational window $\Delta\phi^{k} = \phi^{k}_{max} - \phi^{k}_{min}$.  By definition it gives the possible variation of the azimuthal angle on the photon trajectories starting at a given radial coordinate $r_{source}$, which can be observed at a given inclination angle $i$ and observer position $r_{obs}$ after making $k/2$ loops around the coordinate origin. Then, the solutions of eq. ($\ref{orbit_im}$) for a given boundary data $\{r_{source}, r_{obs}, i\}$ will correspond  to all the possible intersections of the curve $\phi(b)$ with the different observational windows $\Delta\phi^{k}$. From this graphical representation we can deduce  information about the existence of the images of a given order $k$ and some of their qualitative features like the formation of multiple images for example \cite{Nedkova:2020}. Multiple images of order $k$ arise if the intersection of the curve $\phi(b)$ with the observation window $\Delta\phi^{k}$ consists of multiple disconnected portions. Each disconnected piece of the curve $\phi(b)$ leads to a separate image in the observer's sky.

\begin{figure}[t!]
    		\setlength{\tabcolsep}{ 0 pt }{\small\tt
		\begin{tabular}{ cc}
           \includegraphics[width=0.5\textwidth]{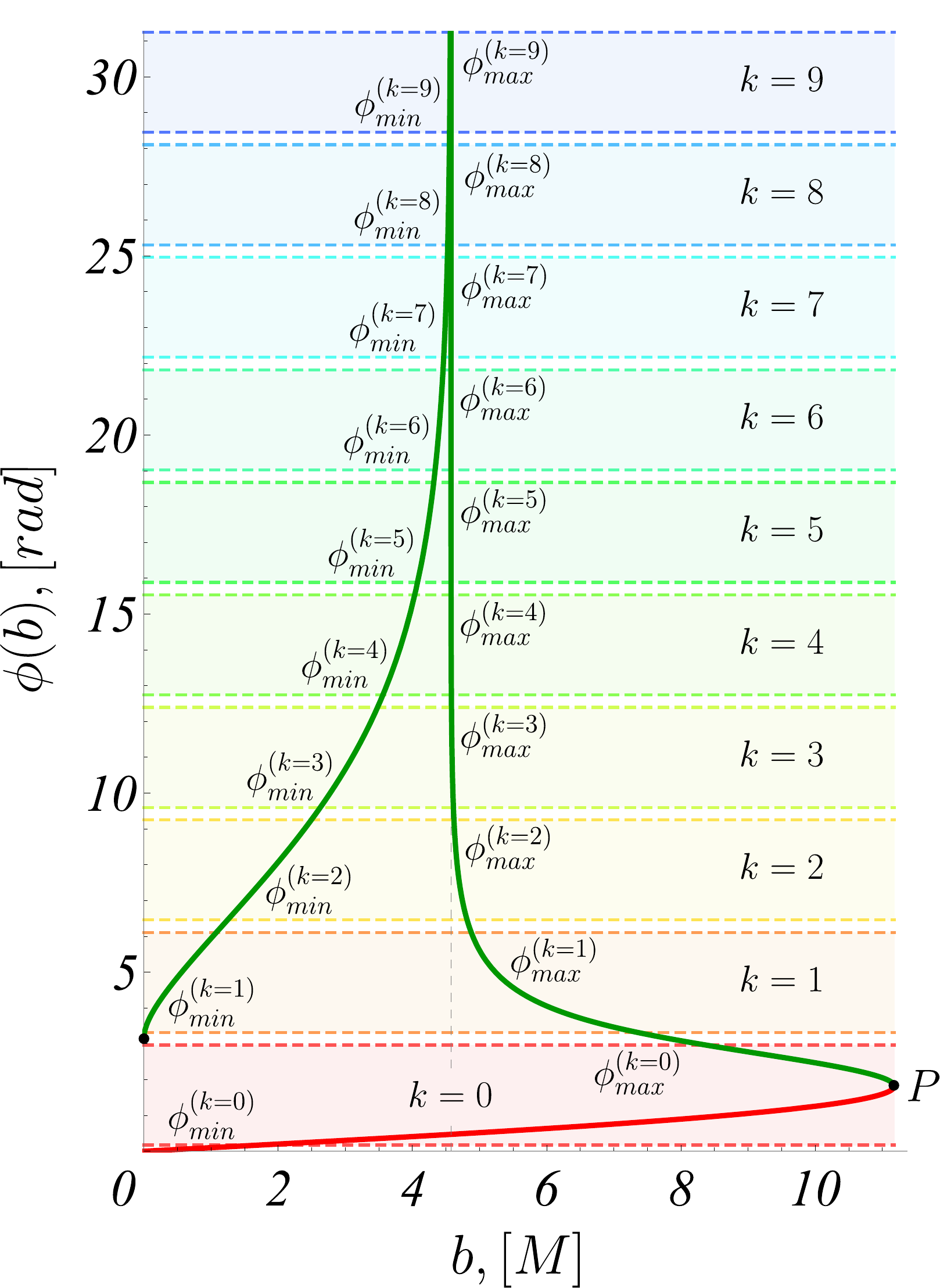}
		   \includegraphics[width=0.5\textwidth]{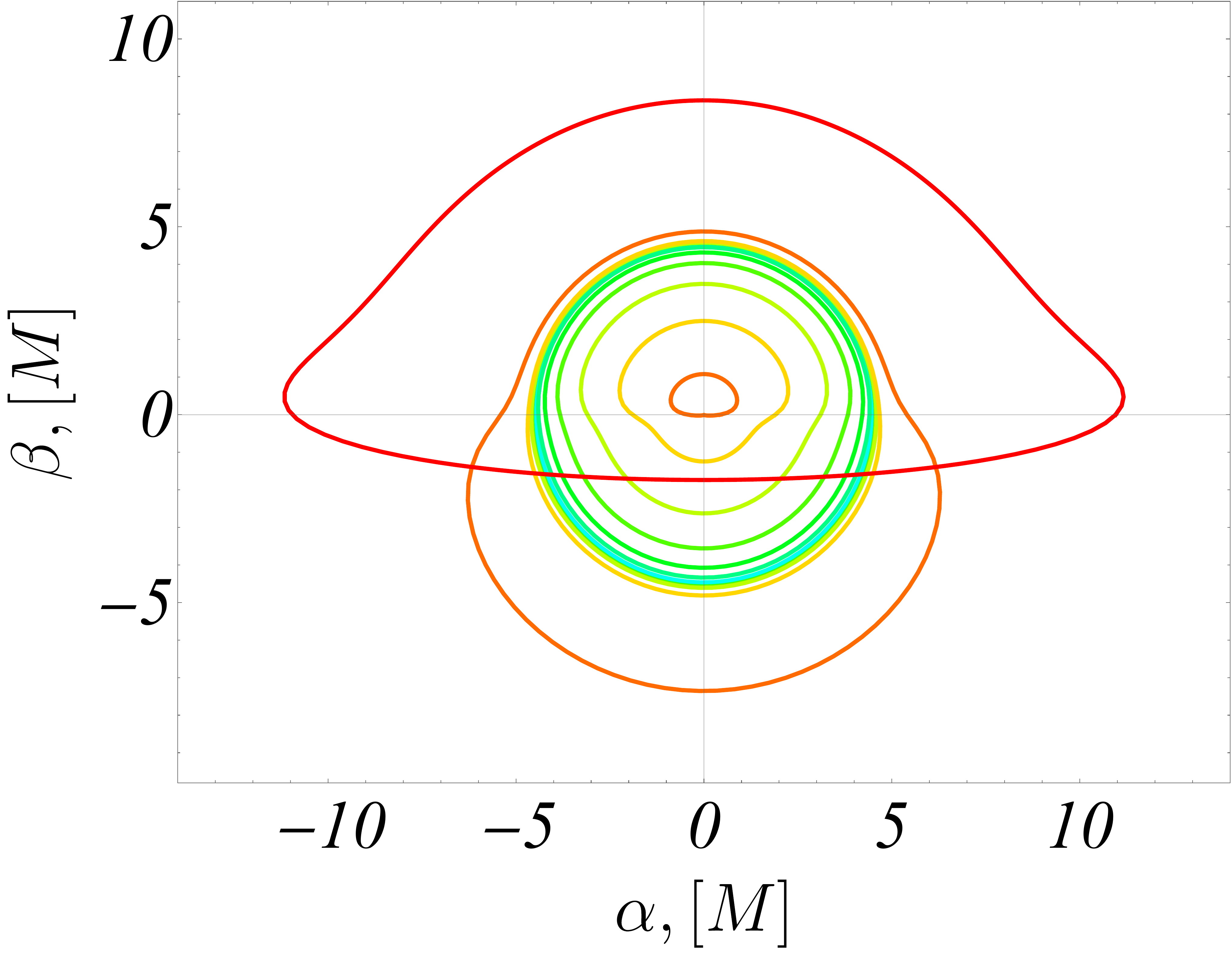} \\[1mm]
           \hspace{0.7cm}  $a)$ \hspace{7.5cm}  $b)$
        \end{tabular}}
 \caption{\label{fig:Image_OD}\small Image formation diagram  for a circular orbit from the outer disk a), and the corresponding observable image b). The orbit is located at the radial coordinate $r=10M$, while the observer's position is at $\{r_{obs} = 5000M, i=80^\circ\}$. In a) we have denoted the value of the impact parameter of the photon sphere with a vertical dashed line.}
\end{figure}

In fig. $\ref{fig:Image_OD}$ we illustrate our argument by presenting the solutions of eq. $(\ref{orbit_im})$ for the boundary data $\{r_{source}=10 M, r_{obs}=5000 M, i=80^\circ\}$. This diagram determines the image of a circular orbit located at $r=10 M$ observed at the inclination angle  $i=80^\circ$.  The red curve represents the solutions of the integral $(\ref{phi_b})$  when the photon trajectories possess no radial turning point, while the green curve corresponds to the solutions when a turning point is present. Although we consider a particular orbit, the properties of its image are representative for any circular orbit from the outer disk.

We see that the graph of the function $\phi(b)$ consists of two disconnected curves, which diverge when $b\rightarrow b_{ph}^{+}$ and $b\rightarrow b_{ph}^{-}$, respectively, where $b_{ph}$ is the value of the impact parameter corresponding to the photon sphere. The curve approaching the photon sphere with values of the impact parameter larger than $b_{ph}$ possesses similar  behavior to the function $\phi(b)$ for the Schwarzschild black hole (see \cite{Nedkova:2020}). For the Schwarzschild black hole eq. ($\ref{orbit_im}$) has a single solution for every $k$, which leads to the formation of an infinite sequence of images approaching the image of the photon sphere when $k\rightarrow\infty$. This reflects the behavior of the trajectories with impact parameters close to $b_{ph}$, since they can perform arbitrary large number of turns abound the photon sphere before reaching the distant observer.

The distinctive feature of the weakly naked Gauss-Bonnet singularity is the existence of a second solution of equation ($\ref{orbit_im}$)  for every $k\geq 1$, forming a second curve  $\phi(b)$, which tends to infinity when approaching the photon sphere, however with values of the impact parameter lower that $b_{ph}$.  Thus, we get a second infinite sequence of images resulting from trajectories which revolve a certain number of times around the photon sphere before scattering to infinity.

\begin{figure}[t!]
\centering
    		\setlength{\tabcolsep}{ 0 pt }{\small\tt
		        \includegraphics[width=0.6\textwidth]{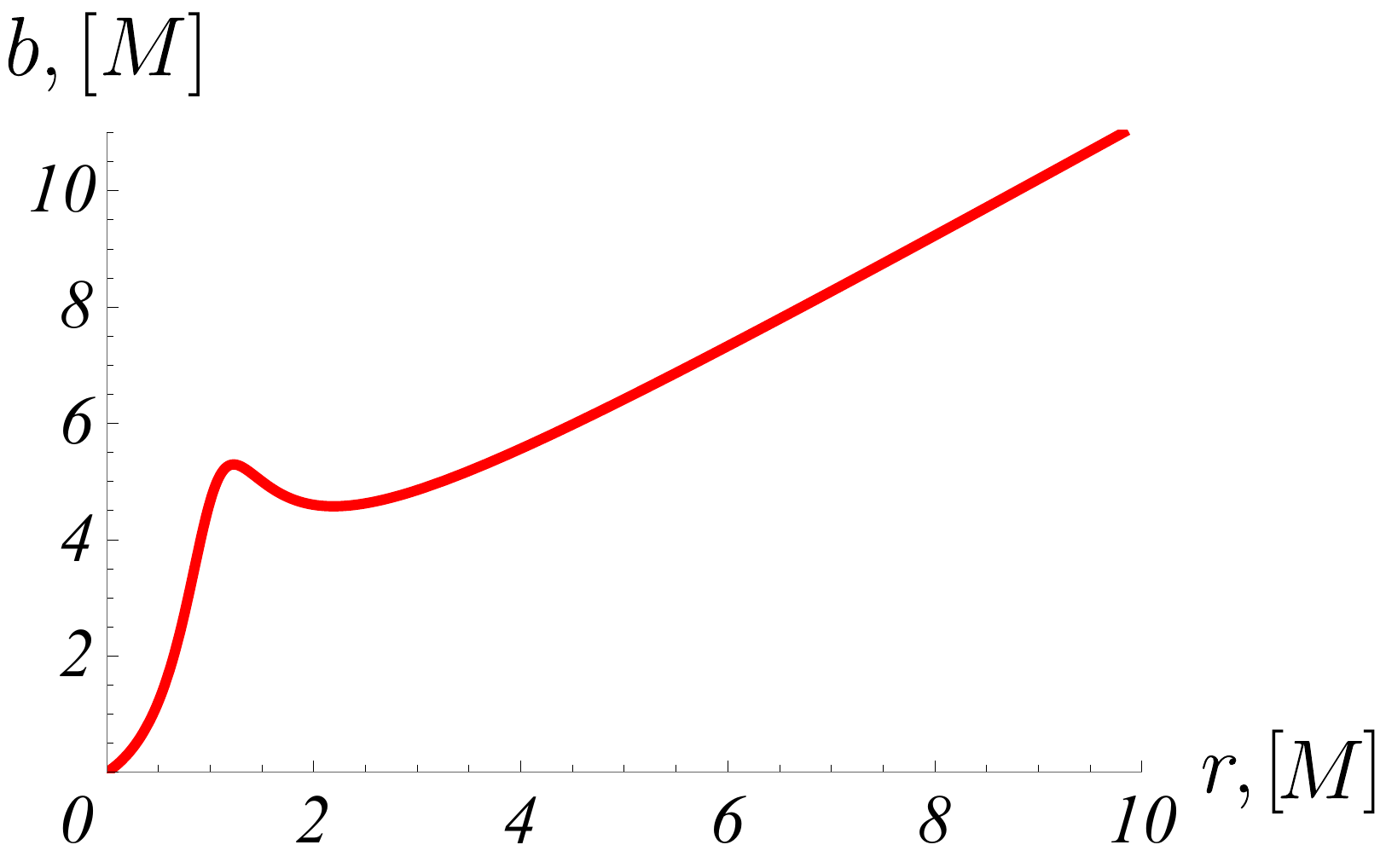}}
 \caption{\label{fig:b_r}\small Dependence of the impact parameter on the photon trajectory's turning point.}
\end{figure}

\begin{figure}[t!]
\centering
    		\setlength{\tabcolsep}{ 0 pt }{\small\tt
		        \includegraphics[width=\textwidth]{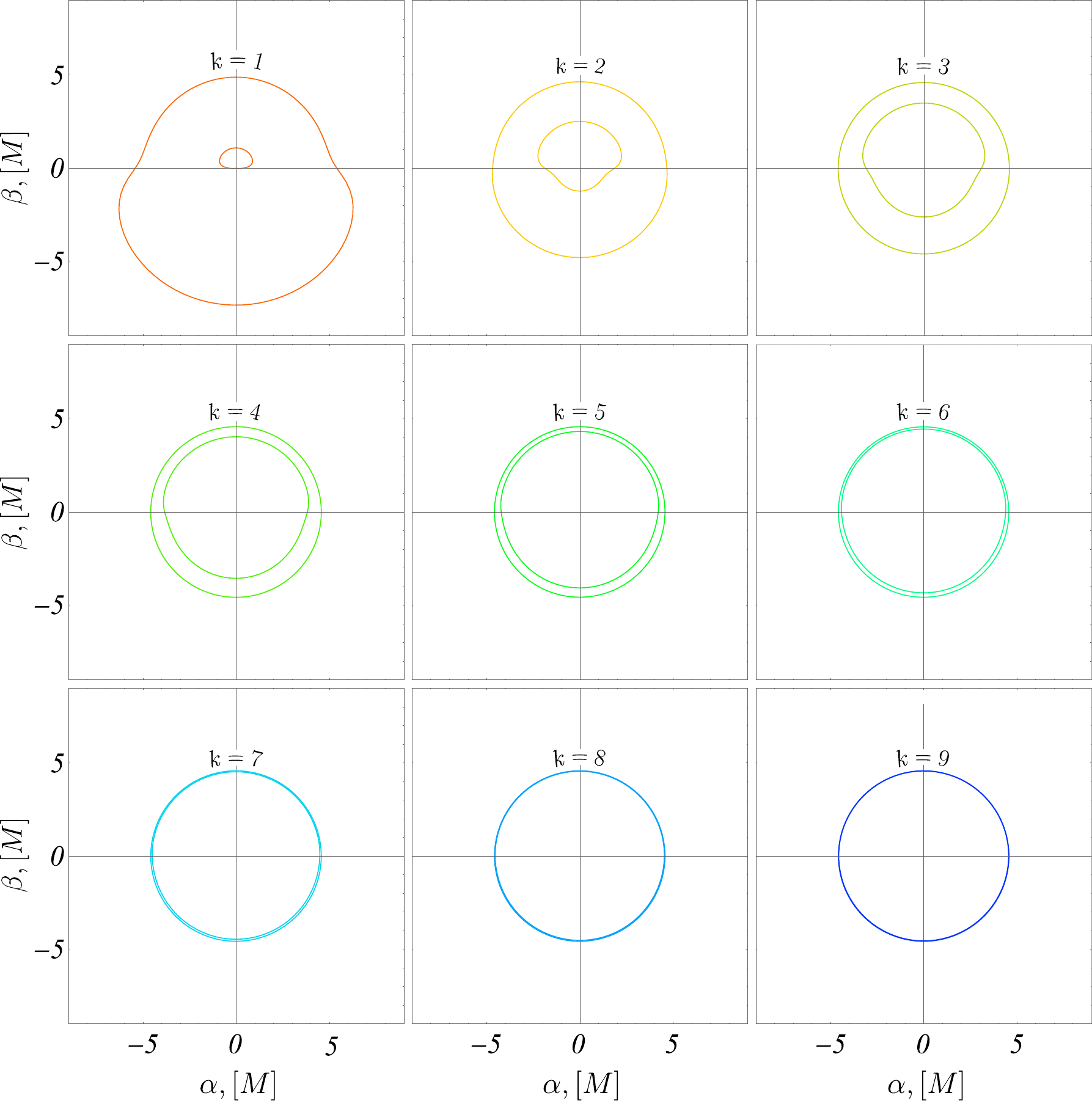}}
 \caption{\label{fig:Image_OD1}\small Images of the circular orbit located at $r=10M$ as seen by an observer at $\{r_{obs}=5000M, i=80^\circ\}$. We illustrate the secondary images up to the order $k=9$.}
\end{figure}

This behavior can be explained by looking at the effective potential for the null geodesics or equivalently at the dependence of the impact parameter on the trajectory's turning point $b(r_0)$ presented in fig. $\ref{fig:b_r}$. The minimum of the curve with coordinates $\{r_{ph}, b_{ph}\}$ corresponds to the photon sphere, while the maximum is located at the stable light ring. We see that there are two classes of photon trajectories which can scatter away to infinity after being emitted at a radial distance larger than the location of the photon sphere. The first type has impact parameters  $b>b_{ph}$, and radial turning points $r>r_{ph}$. This is the only possible case of scattering null geodesics for the Schwarzschild and Gauss-Bonnet black holes, as well as for some other weakly naked singularities like the Janis-Newman-Winicour solution. The second type has impact parameters lower than $b_{ph}$  and turning points approaching the singularity. Such geodesics are possible if the gravitational field becomes repulsive in some characteristic neighbourhood of the compact object and prevents particles and light from reaching it. As a result, all the geodesics emitted in the exterior of the photon sphere will scatter away to infinity either from the first or the second type of potential barriers except for the limit case of trajectories revolving eternally around the photon sphere.

In fig. $\ref{fig:Image_OD}$ b) and fig. $\ref{fig:Image_OD1}$ we present the images of the circular orbit at $r=10M$ up to the order $k=9$ as seen by an observer located at $\{r_{obs} = 5000M, i=80^\circ\}$. The direct image ($k=0$) leads to the characteristic hat-like  shape of the accretion disk primary image observed also for the Schwarzschild and the Gauss-Bonnet black holes at large inclination angles. This is the only order which produces a single image. For all the higher orders $k\geq 1$  a pair of secondary images is formed, as the two images come closer to each other when $k$ increases both approaching the image of the photon sphere in the limit $k\rightarrow\infty$. All the images  of the circular orbit  up to the order $k=9$ are superposed in fig. $\ref{fig:Image_OD}$ b) showing how the image of the accretion disk is built up.

\begin{figure}[t!]
    		\setlength{\tabcolsep}{ 0 pt }{\small\tt
		\begin{tabular}{ cc}
           \includegraphics[width=0.5\textwidth]{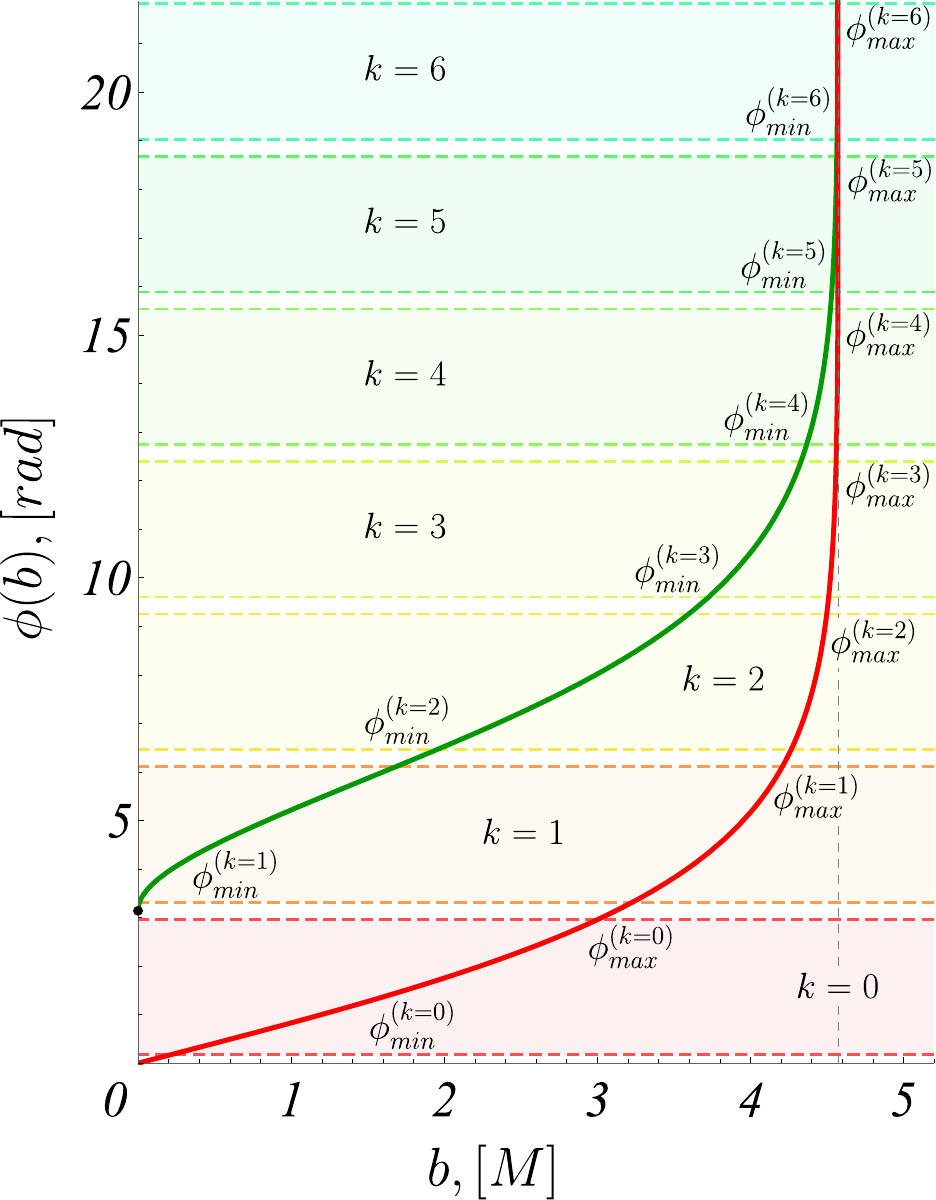}
		   \includegraphics[width=0.5\textwidth]{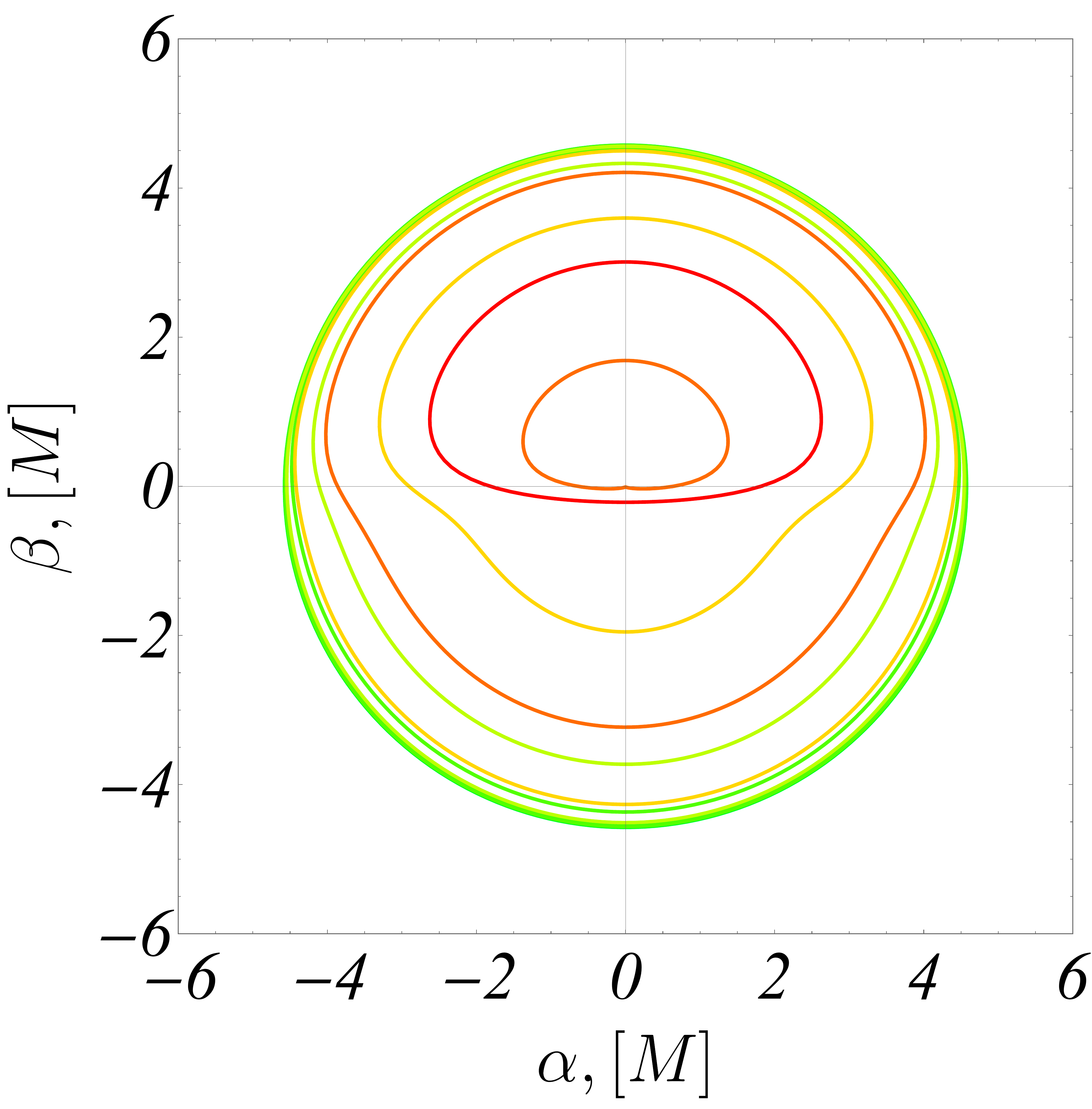} \\[1mm]
           \hspace{0.7cm}  $a)$ \hspace{7.5cm}  $b)$
        \end{tabular}}
 \caption{\label{fig:Image_ID}\small Image formation diagram  for a circular orbit from the inner disk a), and the corresponding observable image b). The orbit is located at the radial coordinate $r=1.227M$, while the observer's position is at $\{r_{obs} = 5000M, i=80^\circ\}$. In a) we have denoted the value of the impact parameter of the photon sphere with a vertical dashed line.}
\end{figure}

\begin{figure}[t!]
\centering
    		\setlength{\tabcolsep}{ 0 pt }{\small\tt
		        \includegraphics[width=\textwidth]{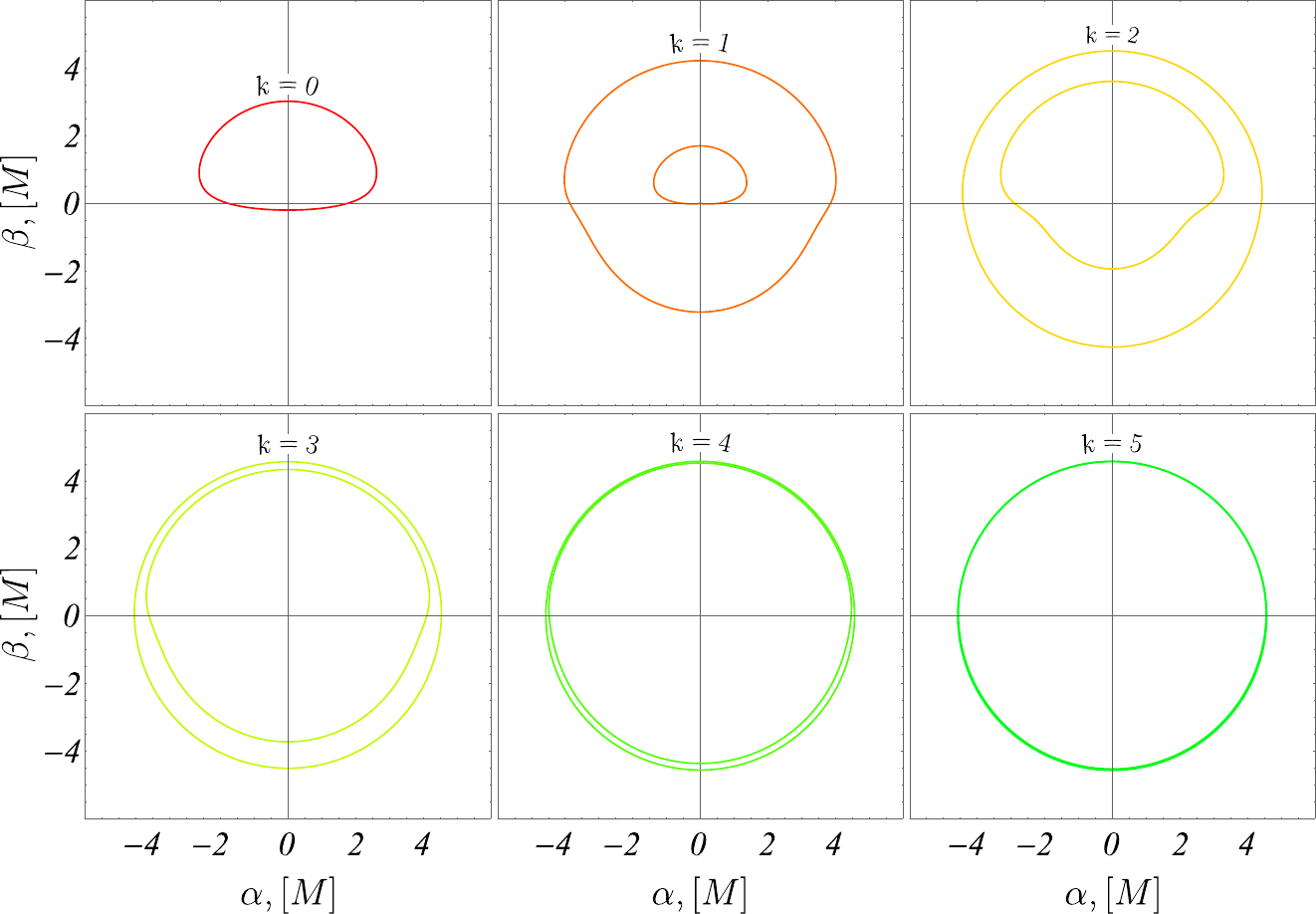}}
 \caption{\label{fig:Image_ID1}\small Images of the circular orbit located at $r=1.227M$ as seen by an observer at $\{r_{obs}=5000M, i=80^\circ\}$. We illustrate the primary image of order $k=0$ and the secondary images up to the order $k=5$.}
\end{figure}

We can use a similar argument to describe the image of the inner disk. The image formation diagram in this case is presented in fig. $\ref{fig:Image_ID}$ for a representative orbit located at $r_{source} =1.227M$, and an observer at $\{r_{obs}=5000M, i = 80^\circ\}$. The images of the orbit up to the order $k=5$ and their superposition are presented in figs. $\ref{fig:Image_ID}$ b) and $\ref{fig:Image_ID1}$.  We observe again a single direct image with $k=0$, while we have a pair of  secondary images for all the higher orders $k\geq 1$. When $k\rightarrow\infty$ the couple of images approach the image of the photon sphere both with impact parameters $b<b_{ph}$. In contrast to the case of the outer disk,  one of the images for each $k$ arises from photon trajectories without turning point, while for the second one we have a turning point. The pattern for the formation of the images can be understood qualitatively by looking at the effective potential for the null geodesics in fig. $\ref{fig:V_eff}$. The photons emitted from the inner disk, i.e. at radial distances smaller than the location of the stable light ring, can reach the spacetime infinity only if their impact parameter is lower than that of the photon sphere. Their trajectories can be either direct, i.e. without a radial turning point, or they can scatter away from the repulsive gravitational field in the vicinity of the singularity. In both cases for suitable impact parameters they can perform arbitrary number of turns around the photon sphere before reaching the observer, thus creating an infinite sequence of higher order images.

\section{Conclusion}

We study the images of the thin accretion disk around static spherically symmetric black holes and  naked singularities in the 4D Einstein-Gauss-Bonnet gravity. On the one hand our goal is to get some insights about the observational signatures of the Gauss-Bonnet gravity compared to the general relativity. In this respect we observe no qualitative distinction in the appearance of the thin accretion disk around the Gauss-Bonnet black holes and the Schwarzschild black hole. Only small quantitative differences are present in the disk size and the maximum of the radiation flux.

On the other hand we aim at investigating the observable features of the naked singularities, which are not present for black holes and could serve as an experimental test for distinguishing the two types of compact objects. For certain values of the coupling constant Gauss-Bonnet gravity allows for an interesting case of naked singularities with respect to their lensing properties. In addition to a photon sphere they possess a stable light ring and the gravitational field becomes repulsive in a certain region around the singularity. Due to these properties broader classes of photon trajectories scatter away from the compact object and reach a distant observer leading to the formation of multiple observable images of the accretion disk. These images appear as a series of bright rings in the center of the primary disk image, and some of them emit radiation  $10^3$ times larger that the flux from the primary disk image. Thus, they represent a significant observational feature. We describe in detail the physical mechanism for the formation of the accretion disk image showing that the phenomenon is expected also for other spacetimes possessing the same properties of the photon dynamics.

\section*{Acknowledgments}
We gratefully acknowledge support by the Bulgarian NSF Grant KP-06-H38/2. Networking support by the COST Actions CA16104 and CA16214 is also gratefully acknowledged.

\end{document}